\documentclass[12pt]{article}
\usepackage{amsmath,amssymb}
\date{} 

\begin{document}
\large

\title{Gamma-ray bursts: a Centauro's cry?}
\author{Z.~K.~Silagadze
\vspace*{3mm} \\
%\address
{Budker Institute of Nuclear Physics,  630 090, Novosibirsk, Russia}
}

\maketitle

\begin{abstract}
Gamma-ray bursts are enigmatic flashes of gamma-rays at cosmological 
distances, so bright that the implied energetics is astounding: energies
of order of about solar rest-energy are liberated in a time scale of the
order of seconds in space regions only a few kilometers in size. Central 
engines capable to produce such enormous explosions, leading to a highly
relativistic expending fireballs, remain a mystery. Here we propose a new
candidate for the gamma-ray bursts central engine.
\end{abstract}
%\PACS{98.70.Rz, 12.39.Fe}

Centauros are unusual cosmic-ray events firstly observed in 1972 at the 
Chacaltaya high mountain laboratory \cite{1}. Detector used in this 
experiment consisted by upper and lower chambers separated by the carbon 
target. Usually the upper chamber detects much more particles than the 
lower chamber, because the electromagnetic component of the shower,
initiated by the primary cosmic-ray interaction at some altitude above the
detector, is strongly suppressed by the carbon layer. The first Centauro
event with the contrary situation came as a big surprise. It had a striking
imbalance between electromagnetic and hadron components and can be 
interpreted as a production of 74 hadrons and only one electromagnetic
($e/\gamma$) particle. Therefore, in Centauro events one can not guess the
lower part of the detector response from the upper one -- hence the name.   

In a sense, gamma-ray bursts are also like Centauros -- from their observed
upper parts it is not easy to guess the inside central engine. Let us see
if we can give some depth to this metaphor.

The clue is provided by one of ``our most perfect physical theories'' 
\cite{2} -- quantum chromodynamics (QCD). At low energies the QCD effective
coupling constant is large and usual perturbation theory methods do not work.
But, fortunately, many features of the low-energy dynamics are dictated by
symmetries of the QCD Lagrangian and their breaking patterns. This enables 
us to substitute QCD at these energies by some effective theory, for example, 
by linear sigma model \cite{3}.

In two flavour case, the Lagrangian of the linear sigma model has the form
\begin{equation}
 L =\frac{1}{2}\,\partial_\mu\sigma \,\partial^\mu\sigma+
\frac{1}{2}\,\partial_\mu\vec{\pi}\cdot \partial^\mu\vec{\pi}-\frac{\lambda}
{4}\left (\sigma^2+\vec{\pi}^{\,2}-v^2\right )^2+H\sigma,
\label{GML} \end{equation}
Here $\lambda\sim 20,\; v\sim 90~\mathrm{MeV}$ and $H\sim (120~\mathrm{MeV})
^3$ are free parameters of the model and they can be fixed by using the pion
and sigma meson masses and PCAC relation as inputs \cite{4} (note that 
\cite{4} is a pedagogical review article where relevant citations to the 
original papers can be found). In the absence
of the last term, the linear sigma model potential has ``Mexican hat'' shape.
Therefore, in this chiral limit the vacuum state is degenerate and chiral
symmetry is spontaneously broken as the sigma field develops a nonzero vacuum
expectation value.

At finite temperature the thermal fluctuations modify the potential \cite{5}.
In a crude, but intuitively transparent approximation the modified potential
takes the form (in the chiral limit. See \cite{4} for explanations)
\begin{equation}
V=\frac{\lambda}{4}\left (\sigma^2+\vec{\pi}^2+\frac{T^2}{4}-v^2
\right )^2. \label{thpot} \end{equation}
The $\sigma$-condensate, which minimizes the energy, now becomes
$$<\sigma>=\sqrt{v^2-\frac{T^2}{4}}$$
and completely melts away at the critical temperature $T_c=2v\approx 
180~{\mathrm MeV}$. Above this phase transition point the chiral symmetry is
restored. Note that this critical temperature is not accurate by about 20-30 
percents. In reality the temperature dependence of the effective potential
is quite complicated - see references cited in \cite{4}. But these subtleties
are completely irrelevant for our purposes - we are just preparing a stage 
for the main idea which will follow.

The last term $L_{SB}=H\sigma$  in the Lagrangian (\ref{GML}) 
explicitly violates the chiral symmetry and originates from the nonzero quark
masses. It tilts the ``Mexican hat'' and removes the vacuum degeneracy. As
a result, $\sigma$-condensate does not melts completely away. However, near
the critical point the residual value of the $\sigma$-condensate is quite 
small and irrelevant \cite{4}.

In the fireball, produced by a primary cosmic-ray particle interaction with
air nuclei, the initial temperature can reach $T_c$. Then the chiral symmetry
is restored and the sigma field condensate melts away 
(up to small corrections due to finite quark masses) in some small volume. But 
the fireball expends quickly and the inside temperature suddenly drops to 
zero. At that the fields do not have enough time to follow this sudden change 
in the environment and, therefore, immediately after the quench the system 
finds itself in a highly out of equilibrium situation for the zero temperature 
Lagrangian -- near the top of the ``Mexican hat''. 

The vacuum expectation values develop while the system begins to roll down
towards the valley of the potential. But all the information about the 
``correct'' orientation of the chiral order parameter (along the sigma 
direction) is lost during the phase transition. Besides, the fireball inside
region is temporally shielded from the outside ``normal'' world by a thin 
shell of hot hadronic debris of the fireball. Therefore we have every reason
to expect that by the time the fields reach the ``Mexican hat'' brim the 
chiral order parameter will be misaligned. Afterwards, then the shielding 
shell disappears, such  Disoriented Chiral Condensate (DCC) relaxes to the 
ordinary vacuum by  emitting a coherent burst of low energy pions 
\cite{4,6,7,8,8a} (the DCC literature is vast, we cite only some 
recent reviews).
 
The formation and decay of the DCC bubble appears to be an 
attractive explanation of the Centauro puzzle, although not without 
difficulties \cite{1,9}. In accelerator based experiments Centauros were
searched but not yet found \cite{1}. These efforts will be continued in the
majority of future heavy-ion experiments and, hopefully, we will know soon 
whether the DCC phenomenon really exists.

Now let us return to the gamma-ray bursts \cite{10,11,12,13} and put our main 
card on the table: if a macroscopic DCC region might be formed in some 
violent cosmic event, its subsequent decay will provide a highly efficient 
source of a pure radiation energy without the baryon loading problem.

Observational evidence suggests that at least long gamma-ray bursts are 
associated with star-forming regions and thus most probably with the death
(collapse) of massive stars \cite{14}. During supernova core collapse, 
temperatures can be as high as about 40~MeV \cite{15}, but this is still much
lower than the critical temperature $T_c$ discussed above. How can the
desired chiral phase transition then happen?

The value of $T_c$, extracted from the simple arguments concerning 
high-tempe\-ra\-ture behaviour of the linear $\sigma$-model, 
assumes low density 
environment, while in supernova core the density is high. Therefore, in 
contrast to the cosmic-ray and heavy-ion experiments, we need high density 
QCD, not the high temperature one.
 
At high density, one also expects the restoration of chiral symmetry. This is
especially transparent in the MIT bag-model picture of nucleons as 
droplets of chiral symmetry restored phase, within which the quark density
is nonzero, immersed in a sea of nonperturbative vacuum with no quarks but
nonzero chiral condensate. While squeezing a lump of hadronic matter, as the 
outer pressure increases, the chiral symmetry restored phase occupies more 
and more fraction of the volume. At last individual bags begin to overlap
and finally they unify in a one big bag containing all the quarks. In such so
called quark-gluon plasma (QGP) phase, the chiral symmetry is restored up to
effects of nonzero quark masses. 

The real situation is not as simple, however. The phase structure of 
high-density QCD turns out to be surprisingly reach and its study is an 
active research field at present \cite{16,17,18,19}. A main result is that
at high-densities one expects formation of diquark condensates and
development of colour superconductivity, due to Fermi surface generic 
instability against attractive interactions.

In the colour superconductive phase, the chiral symmetry can be both restored,
as in the so called two flavour colour superconductivity ($2SC$) phase, and 
spontaneously violated, as in the colour-flavour-locked ($CFL$) phase. 
In latter case, however, the chiral symmetry breaking mechanism is different
from the ordinary one at low densities. It turns out that in the $CFL$ phase 
some discrete chiral symmetry remains  unbroken up to small instanton-induced
effects \cite{20,21}. This residual (approximate) symmetry  renders difficult
the development of ordinary $<\bar q_L q_R>$ chiral condensate and as a 
result its magnitude is much less than in the ordinary QCD vacuum \cite{21a}.
  
Therefore, it seems  plausible that the transient phase transition, triggered
by a high-density, can also lead to formation of DCC bubbles. Let us take 
a closer look at some speculative scenarios how this may happen.

Interesting and still open fascinating possibility is the hypothetic stability
of three-flavour quark matter (strange matter) \cite{22,23}. If the strange 
matter is indeed the true ground state of hadronic matter, instead of 
$\;^{56}Fe$, then strange stars may exist \cite{24}. Neutron star ($NS$) 
will be converted into strange star ($SS$), if a strange matter seed is 
created inside the neutron star, big enough to overcome surface energy cost. 
In fact, $NS\to SS$ conversion was suggested as a candidate for the gamma-ray 
bursts central engine \cite{25,26,27,28,29,30} (we cite only some relevant 
references, others can be traced through them).

A newborn neutron star can increase its central density because of accretion
or spin-down. When the density becomes sufficiently high, the deconfinement
phase transition happens and two-flavour ($u$ and $d$) quark matter is formed
in the inner core \cite{24a}. But the two-flavour quark matter is unstable 
and converts
into three-flavour strange matter via weak interactions very rapidly, with 
a time scale below $10^{-7}~\mathrm{s}$ \cite{31}. Once a strange matter seed
is formed, it will start to grow and swallow the surrounding neutron matter.
Combustion is limited by quark diffusion and might be, in principle, slow
\cite{32}, if not the hydrodynamical instabilities \cite{33} which, most 
probably, will turn the conversion of neutron matter into detonation. The
whole star will be digested in a time period maybe as short as 
$10^{-3}~\mathrm{s}$ \cite{28}.  

The $NS\to SS$ conversion is accompanied by a release of huge amount of 
internal energy, comparable to the star's binding energy. For example, 
according to estimates \cite{28}, a neutron star with a mass $M_{NS}=
1.409\,M_\odot$ and a radius $R_{NS}=11.0~\mathrm{km}$, which has 
a gravitational binding energy of about $4.5\times 10^{53}~\mathrm{erg}$,
after the conversion produces the strange star with $M_{SS}=1.254\,M_\odot$ 
and $R_{SS}=10.5~\mathrm{km}$. At that the expected amount of internal energy
released is $(1.9-4.2)\times 10^{53}~\mathrm{erg}$. As these estimates show,
the liberated internal energy can constitute a significant fraction of the
gravitational binding energy. Therefore a nascent strange matter will inflate
somewhat before it reassembles itself into the strange star. Hence the 
appearance of vacuum bubbles inside strange matter seems inevitable. But which
vacuum? The primordial chiral condensate disappears during the deconfinement 
phase transition and the consequent neutron matter burning into strange 
matter. The new vacuum inside the emergent bubbles is shielded for some time 
from the normal vacuum by the strange matter. Therefore, it seems natural to 
expect that the QCD vacuum will be disoriented inside these bubbles, that is 
DCC will be formed. Afterwards, when the DCC bubbles will come into contact 
with the ordinary vacuum, they will decay by emitting coherent pions.

Let us estimate how much energy can be stored in these DCC bubbles. Suppose 
DCC domain is formed inside the bubble with the misalignment angle
$\theta$. That is inside this DCC region one has
$$<\sigma>_{DCC}=f_\pi\cos{\theta},\;\;\; <\vec{\pi}>_{DCC}=f_\pi\sin{\theta}
~\vec{n},$$
where $\vec{n}$ is a unit vector in isospin space and 
$f_\pi\approx 93~\mathrm{MeV}$ is the pion decay constant. While in the normal 
vacuum 
$$<\sigma>=f_\pi,\;\;\;<\vec{\pi}>=0.$$
Energy density in the DCC  region is higher than in the normal vacuum because 
of the symmetry breaking term $V_{SB}=-H\sigma$. The difference equals (note
that $H=f_\pi m_\pi^2$)
$$\Delta \epsilon=-H<\sigma>_{DCC}+H<\sigma>=
Hf_\pi(1-\cos{\theta})=2f_\pi^2 m_\pi^2\sin^2{\frac{\theta}{2}}\;.$$
As an estimator for the total volume of the produced DCC, we take the volume 
of the spherical shell with inner radius $R_{SS}$ and outer radius $R_{NS}$.
Besides, $\sin^2{\frac{\theta}{2}}$ can be replaced by its mean value $0.5$,
because the misalignment angle $\theta$ changes randomly from domain to 
domain in the bubbles. Then the total energy content of the DCC bubbles equals
\begin{equation}
E_{DCC}=\frac{4\pi}{3}f_\pi^2m_\pi^2(R_{NS}^3-R_{SS}^3)\approx
\frac{R_{NS}^3-R_{SS}^3}{1~\mathrm{km}^3}~1.5\times 10^{50}~\mathrm{erg}.
\label{EDCC} \end{equation}
For $R_{NS}=11.0~\mathrm{km}$ and $R_{SS}=10.5~\mathrm{km}$ this estimation 
results in $E_{DCC}\approx 2.6\times 10^{52}~\mathrm{erg}$ -- just right
amount of energy to power a gamma-ray burst!

One can imagine another, even more speculative, scenario which could lead to
formation of macroscopic DCC regions. The density in a supernova core can be
raised to extreme values during the catastrophic gravitational collapse. It
was argued \cite{34} that even the electroweak symmetry has a chance to be
restored in such collapse before the corresponding region is engulfed by a
emergent black hole horizon. The subsequent baryon burning due to the baryon
number violating processes, unsuppressed in the symmetric phase, was 
suggested as one more candidate for the gamma-ray bursts central engine
\cite{34}.   

The chiral symmetry will be restored much before the electroweak symmetry.
However, to form the DCC regions one needs gravity suddenly to cease its
deadly grasp of these chirally symmetric areas. Otherwise they will find 
their way into the central singularity with final fate obscure at present. 
Clearly, the latter is the only possibility offered by the classical theory 
of gravity.

However, a challenging possibility, that General Relativity is just a low 
energy effective theory \cite{35,36}, can not be excluded. Therefore, the 
gravitational collapse may have completely different outcome than it is
expected in Einstein gravity. The following analogy shows this picturesquely
\cite{36}. A stretched rubber sheet with a heavy ball on it models general
relativity. The membrane is an analog of space-time, the ball corresponds 
to some gravitating object and the distortion of the rubber sheet surface
represents the gravitational field. By declaring the laws of elasticity to be
universally true, no matter how extreme the stretching, we expect that when 
the weight of the ball increases, so does the depression making small
objects, which fall into this distortion, more and more inaccessible for us.
In reality, however, the membrane ruptures if the ball becomes too heavy and
the objects fall through.

In the condensed matter analogs of gravity, the vacuum state also does not 
remain stable in the presence of the horizon \cite{37}. Besides, perturbations
of the quantum vacuum, for example, due to QCD or electroweak phase 
transitions, can alter the cosmological constant, which in the effective 
gravity is not a constant but the evolving physical parameter \cite{37}.  
Therefore, one can not exclude that the gravitational collapse will be
accompanied by much reacher phenomenology, like formation of DCC, or even more
exotic false vacuum bubbles, than it is anticipated at present.

Although the effective gravity idea \cite{35,36,37,38,39} is very attractive,
the above mentioned possibility of DCC formation during a gravitational 
collapse is much more speculative than the scenario associated
with the $NS\to SS$ transition. Therefore, let us summarize the latter:
\begin{itemize}
\item  $NS\to SS$ transition happens probably very quickly due to detonation. 
In almost all volume, occupied by the initial $NS$, the strange matter is 
formed. But the strange matter has a higher density than the neutron 
matter. Therefore the produced strange matter can not be completely 
continuous, it will have voids (vacuum bubbles) in its body.
\item  The voids will be formed also because the energy liberated 
during the $NS\to SS$ transition is very high: one expects the newborn 
strange matter to inflate somewhat before its lumps reunify and form the 
Strange Star.
\item What we are interested in is the content of these voids. In usual QCD
vacuum one has a nonzero chiral condensate $<\bar q_R q_L>$ -- in the linear
sigma model language the corresponding quantity is a nonzero vacuum 
expectation value of the sigma field.
\item But during deconfinement phase transition this chiral condensate 
disappears. Besides in various forms of quark matter this condensate is 
either absent ($2SC$), or is much smaller ($CFL$) compared to its value in the 
normal QCD vacuum at small density and temperature. So one can expect that 
just after the void bubble is formed inside the strange matter, the fields 
inside the bubble, both pions and sigma, do not have any vacuum expectation
values, or these values are small.
\item This situation is not stable inside the bubble, where we have a small
($\sim$zero) density and a small temperature environment, so that the fields
dynamics is approximately governed by the zero temperature and density 
linear sigma model Lagrangian. The potential of this Lagrangian has a 
"Mexican hat" shape and the initial (small) values of the fields are located 
near its top. So the fields begin to roll down.
\item When the fields reach the "Mexican hat" brim they can find themselves 
in a wrong place (from the point of view of the normal QCD vacuum). So the 
so called disoriented chiral condensate (DCC) may be formed.
\item When the DCC bubble comes into contact to the normal QCD vacuum, it 
will decay by emitting pions. At that, if the DCC was oriented in the $\pi^0$ 
direction (in isospace), it will decay by emitting only neutral pions. If 
the initial orientation was perpendicular to the $\pi^0$ direction, the 
neutral  pions will not appear at all, but one will have equal numbers of 
$\pi^-$ and $\pi^+$.
\item Strong electric and magnetic fields can prefer the DCC oriented in the
$\pi^0$ direction due to chiral anomaly  \cite{40}.
\item  A huge amount of produced $\pi^0$-s will create a wonderful relativistic
fireball because of the decay $\pi^0\to 2\gamma$ (or such a fireball will be 
formed also due to annihilation of $\pi^-$ and $\pi^+$ pairs, if DCC not in 
all bubbles is oriented in the $\pi^0$ direction.
\end{itemize}

In fact the $NS\to SS$ transition is just one example which may lead to the 
formation of macroscopic amount of DCC bubbles. The main idea of the 
paper is more general: if in some energetic cosmic events a macroscopic 
amount of DCC bubbles can be formed, then we will have an effective GRB 
central engine at hand. 

Physics is an experimental science. Therefore, the idea that the cosmic-ray
Centauros and gamma-ray bursts both have a common QCD origin, however 
attractive, must confront scrutiny of the future heavy-ion and cosmic-ray 
experiments, as well as astrophysical observations. Remarkably, more than 
135 theoretical models of the gamma-ray bursts were instantaneously destructed
when the observational data from the Beppo-SAX satellite became available 
\cite{41}. We hope that the idea presented in this article will be more 
fortunate.

\end{document}